\documentclass[aps,prb,twocolumn,psfig,groupedaddress]{revtex4-1}
\usepackage{times}
\usepackage{graphicx,epsf}
\usepackage{bm}
\usepackage{multirow}
\usepackage{subfigure}

\newcommand{\abs}[1]{\left\lvert #1\right\rvert}

\newcounter{lastnote}

\def\bc{\begin{center}}
\def\ec{\end{center}}
\def\be{\begin{equation}}
\def\ee{\end{equation}}

\usepackage{amsmath,amssymb}

\begin{document}
\title{Evidence for electron-electron interaction in topological insulator thin films}
\author{Jian Wang,$^{*1,4}$ Ashley M. DaSilva,$^{*1}$ Cui-Zu Chang,$^{2,3}$ Ke He,$^{2}$  J. K. Jain,$^1$ Nitin Samarth,$^1$ Xu-Cun Ma,$^{2}$  Qi-Kun Xue,$^{3}$ Moses H. W. Chan$^1$}
\affiliation{$^1$The Center for Nanoscale Science and Department of Physics, The Pennsylvania State University, University Park, Pennsylvania 16802-6300, USA}
\affiliation{$^2$Institute of Physics, Chinese Academy of Sciences, Beijing 100190, China}
\affiliation{$^3$Department of Physics, Tsinghua University, Beijing 100084, China}
\affiliation{$^4$International Center for Quantum Materials and State Key Laboratory for Mesoscopic Physics, School of Physics, Peking University, Beijing, 100871, China}
\begin{abstract}
We consider in our work  single crystal thin films of Bi$_{2}$Se$_{3}$, grown by molecular beam epitaxy, both with and without Pb doping. Angle-resolved photoemission data demonstrate topological surface states with a Fermi level lying inside the bulk band gap in the Pb doped films. Transport data show weak localization behavior, as expected for a thin film in the two-dimensional limit (when the thickness is smaller than the inelastic mean free path), but a detailed analysis within the standard theoretical framework of diffusive transport shows that the temperature and magnetic field dependences of resistance cannot be reconciled in a theory that neglects inter-electron interactions. We demonstrate that an excellent account of quantum corrections to conductivity is achieved when both disorder and interaction are taken into account. These results clearly demonstrate that it is crucial to include electron electron interaction for a comprehensive understanding of diffusive transport in topological insulators. While both the ordinary bulk and the topological surface states presumably participate in transport, our analysis does not allow a clear separation of the two contributions.
\end{abstract}

\maketitle

\section{Introduction}

Topological structures in condensed matter often result from interactions; examples include vortices in superconductors, superfluids, two-dimensional XY systems, and the fractional quantum Hall effect. Topology also governs the behavior of certain non-interacting systems:  the integer quantum Hall effect is the quintessential example, while the recently proposed topological band insulators provide another. The latter are predicted to contain conducting surface states with an odd number of Dirac cones that are topologically protected against time reversal invariant perturbation.~\cite{Fu2007, Moore2007, Roy2009, Qi2008} It is important to demonstrate that the topological features of these new systems are robust to interparticle interaction. The existence of surface states has been investigated and confirmed for several candidate topological insulators using angle-resolved photoemission spectroscopy (ARPES).~\cite{Hsieh2009,Hasan2010} Detecting these surface states by transport, which might at first appear the most natural probe, has proved more challenging, largely because of non-negligible bulk conduction. Here, we report a detailed study of the temperature and magnetic field dependence of the conductivity in topological insulator thin films grown by molecular beam epitaxy. A theoretical analysis of the observed quantum corrections to the conductivity unambiguously demonstrates that electron-electron interactions play a crucial role in determining the diffusive transport behavior of topological insulators.

\begin{figure}
\includegraphics[width=.75\columnwidth]{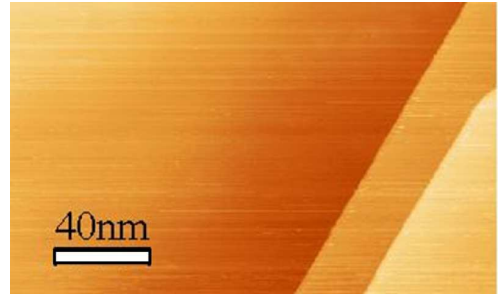} 
\caption{A Scanning Tunneling Microscope (STM) Image of a Typically Grown Bi$_{2}$Se$_{3}$ Film with a Thickness of 45 Quintuple Layers (QLs).}
\label{STM}
\end{figure}

\begin{figure}
\includegraphics[width=.75\columnwidth]{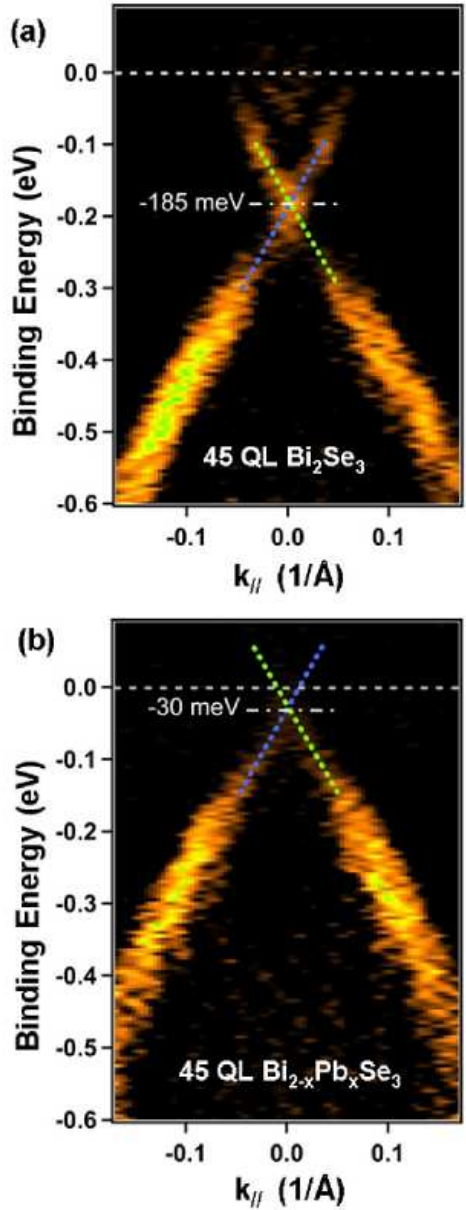} 
\caption{ARPES spectra of the 45 nm thick crystalline films. (a) The undoped Bi$_{2}$Se$_{3}$ film and (b) doped Bi$_{2-x}$Pb$_{x}$Se$_{3}$ film. The energies are measured relative to the Fermi level.}
\label{ARPES}
\end{figure}

Bi$_{2}$Se$_{3}$ is a candidate for a three dimensional topological insulator (TI), provided its Fermi level can be shifted to lie inside the bulk energy gap, as can be accomplished by appropriate amount of doping. Due to its relatively large band gap (0.3 eV) and its simple surface state structure (a single Dirac cone), it has become the canonical reference material for the 3D TIs.~\cite{Zhang2009,Hasan2010, Qi2010} Several transport experiments have suggested the presence of conducting surface states in candidate 3D TIs. 
A large magnetofingerprint signal\cite{Checkelsky2009} 
in macroscopic crystals of nonmetallic Bi$_{2}$Se$_{3}$,
Aharonov-Bohm oscillations in Bi$_{2}$Se$_{3}$ nanoribbons\cite{Peng2010} with dimensions of $\sim$ 100 nm,
and Shubnikov-de Haas (SdH) oscillations\cite{Qu2010} 
in bulk TI  (Bi$_{2}$Te$_{3}$) have been reported and associated with the surface states.
However, another SdH measurement on high quality Bi$_{2}$Se$_{3}$ single crystals did not find significant contributions from surface states and suggested that these are subject to strong scattering.~\cite{Butch2010} It is noted that the analysis of all the above experiments neglects interactions. 

Recent progress in thin film growth of TI materials by molecular beam epitaxy (MBE)~\cite{Cheng2010, Zhang2010} has enabled the fabrication of  thin films of crystalline Bi$_2$Se$_3$.  We study the temperature and magnetic field dependence of two thin films of Bi$_2$Se$_3$, one undoped and the other doped. The latter is doped with an appropriate concentration of Pb, which should ideally render the bulk insulating.  ARPES measurements demonstrate that the latter thin film possesses Dirac-cone surface states with the Fermi level lying in the bulk band gap. Although it is likely that there is bulk conduction in the doped thin film as well, the expectation is that the contribution from surface states will make a significant contribution to the electrical transport. The conductance shows a logarithmic temperature dependence at low temperatures, a hallmark of weak localization (WL) in two dimensions (2D), originating from the interplay between disorder, spin-orbit coupling and/or electron-electron interactions (EEI)\cite{Lee1985}. We find that the temperature and magnetic field dependences of our data are qualitatively inconsistent with conventional 2D transport theory of non-interacting electrons, but are successfully explained when we include, following the theory of Lee and Ramakrishnan~\cite{Lee1982}, quantum corrections to the conductivity originating from EEI and also account for the Zeeman splitting, which is large in these narrow band gap materials~\cite{Kohler1975}. 

\section{Experimental results}

We use MBE to grow 45 QL Bi$_2$Se$_3$ films on bare insulating 6H-SiC (0001) substrates whose resistivity is $\sim 1\times10^6$ $\Omega\cdot$cm, allowing us to neglect the conductance contribution from the substrate. 
A scanning tunneling microscope (STM) image of a typical film is shown in Figure~\ref{STM}. Atomically flat terraces with widths over 100 nm can be observed, with 1 quintuple layer thick terraces. 
Figure~\ref{ARPES}(a) shows the ARPES band map of a 45 QL Bi$_2$Se$_3$ film. A single Dirac cone is observed at the $\bar{\Gamma}$ point, with the Dirac point located at 0.185 eV below the Fermi level. Hence, both the surface states and the bulk conduction band can contribute to electronic transport. The bulk chemical potential can be tuned inside the bulk energy gap by using appropriate dopants.~\cite{Checkelsky2009} In this experiment, we accomplish this by doping the sample with $~ .37 \%$ Pb.~\cite{Zhang2010b} As shown in Fig.~\ref{ARPES}(b), after doping, the Fermi level of the Bi$_{2-x}$Pb$_x$Se$_3$ film is inside the energy gap. In this situation, the bulk conductivity is suppressed and the surface conductance becomes more evident. 

The Bi$_2$Se$_3$ films were grown under Se-rich conditions on 6H-SiC (0001) substrates at 220$^{\circ}$C in an ultrahigh-vacuum (UHV) system (Omicron), equipped with MBE, STM and ARPES~\cite{Zhang2010}. The base pressure of the system is $1.5\times 10^{-10}$ Torr. High purity Bi (99.9999\%) and Se (99.999\%) were thermally evaporated from standard Knudsen cells. The temperatures of the Bi source and Se source are 550 and 170$^{\circ}$C. The Se$_{4}$(Se$_{2}$)/Bi flux ratio was between 10 and 15, which leads to a growth rate of $\sim 0.3$ QL min$^{-1}$. In ARPES measurement, He-I$\alpha$ (21.21 eV) photons produced by a Gammadata VUV 5000 discharging lamp and a Scienta SES2002 analyzer are used to excite and collect photoelectrons, respectively. All STM and ARPES data are taken at room temperature.

Subsequent to the ARPES analysis, the films were covered by 30 nm thick amorphous Se, which is quite insulating, as protective layers. Then, the samples were taken out from the UHV system for transport measurements in a variable temperature magnetocryostat with high magnetic field ($H \leq 80$ kOe) and low temperature ($0.1$ K $\leq T \leq 300$ K) capability. Standard four probe electrical transport measurements were carried out using an ac resistance bridge with an excitation current of 500 nA. All resistance values were obtained by averaging over 50 measurements. 

\begin{figure}
\includegraphics[width=.75\columnwidth]{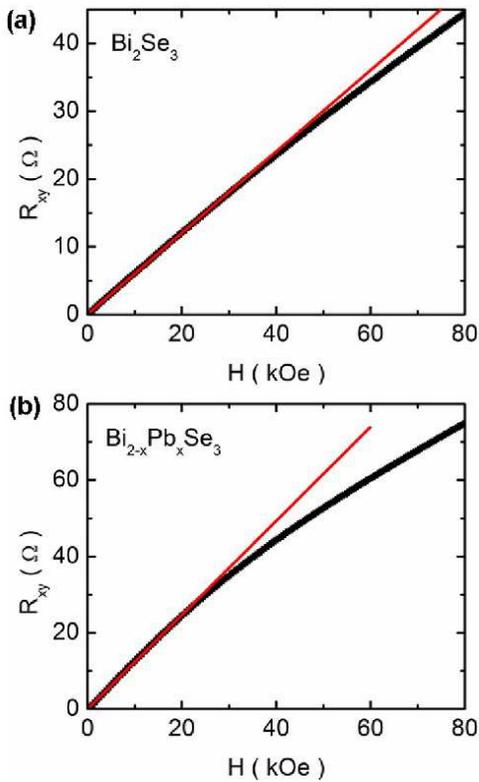}
\caption{Hall Resistance of the 45 QL Bi$_{2}$Se$_{3}$ and Bi$_{2-x}$Pb$_{x}$Se$_{3}$ Films at 2 K. Below 20 kOe, the Hall resistance shows almost linear behavior (the red solid lines in the figure) in both samples. However, with the increasing magnetic field, we can clearly see the deviation from the linear behavior, which is larger in the Bi$_{2-x}$Pb$_{x}$Se$_{3}$ film. The nonlinear Hall behavior is reminiscent of the two channels of carriers in the films. One channel may be from the surface state.}
\label{HR}
\end{figure}

We determine the carrier density in the two samples using the Hall effect, as shown in Fig.~\ref{HR}. The carriers are electrons in both samples, and from the approximately linear behavior of $R_{xy}$ at low magnetic fields, we estimate a carrier density of $2.27\times 10^{19}$~cm$^{-3}$ ($1.02\times 10^{14}$~cm$^{-2}$) for the Bi$_{2}$Se$_{3}$ film and $1.1\times 10^{19}$~cm$^{-3}$ ($4.95\times 10^{13}$~cm$^{-2}$) for the Bi$_{2-x}$Pb$_{x}$Se$_{3}$ film at 2 K. We estimate $k_{F}\ell_{e}\sim 22$ for Bi$_{2}$Se$_{3}$ and $\sim 5.7$ for Bi$_{2-x}$Pb$_{x}$Se$_{3}$, well within the diffusive transport regime. (The quantity $\ell_e$ is the elastic scattering length.) 
$R_{xy}$ has a slightly non-linear behavior at high magnetic fields, which appears to be quite generic at low temperatures, and is likely an indication of more than one transport channel.~\cite{Kong2010,Steinberg2010}

\begin{figure}
\includegraphics[width=.75\columnwidth]{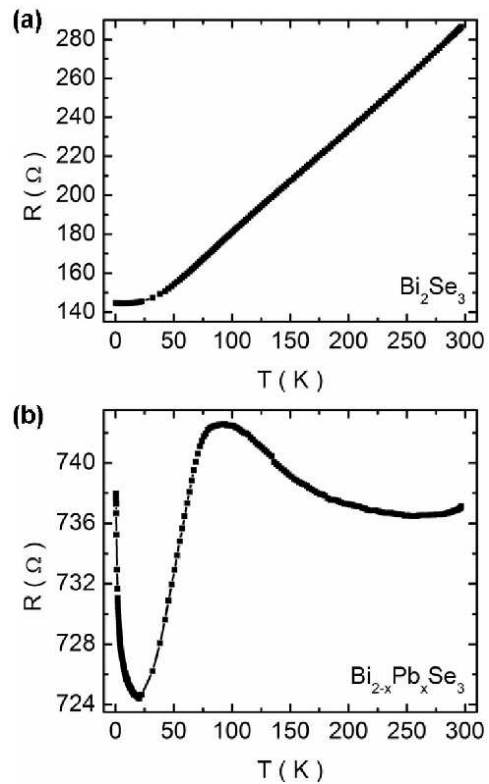}
\caption{Resistance vs Temperature Plots of the 45 QL Bi$_{2}$Se$_{3}$ and Bi$_{2-x}$Pb$_{x}$Se$_{3}$ Films from 300 K to 0.5 K. }
\label{RT}
\end{figure}

The temperature dependence of the sample resistivity in the range $0.1$ K $\leq T \leq 300$ K reveals a qualitative difference between the electrical properties of the two thin films. See Fig.~\ref{RT} for the resistance vs temperature (R-T) plots for the two films from room temperature to 0.5 K. Above 50 K, the resistance of the Bi$_{2}$Se$_{3}$ film increases linearly with temperature, characteristic of the metallic behavior expected for a degenerately doped semiconductor. Its resistivity varies from 1.43 m$\Omega$ cm at 297 K to 0.72 m$\Omega$ cm at 500 mK. For the Bi$_{2-x}$Pb$_{x}$Se$_{3}$ film, the resistance is higher, with only a very weak, non-monotonic T dependence. Its resistivity is 3.69 m$\Omega$ cm at both 297 K and 500 mK. 

In an earlier study\cite{Butch2010} of a bulk Bi$_2$Se$_3$ system, it was found that the bulk remains conductive even for densities much smaller than those quoted above. The origin of the qualitative difference between our two thin films is not understood, but presumably has to do with the low-dimensional nature of our system, and also the fact that it has a much lower mobility than the system in Ref.~\cite{Butch2010}. 

\begin{figure}
\includegraphics[width=.75\columnwidth]{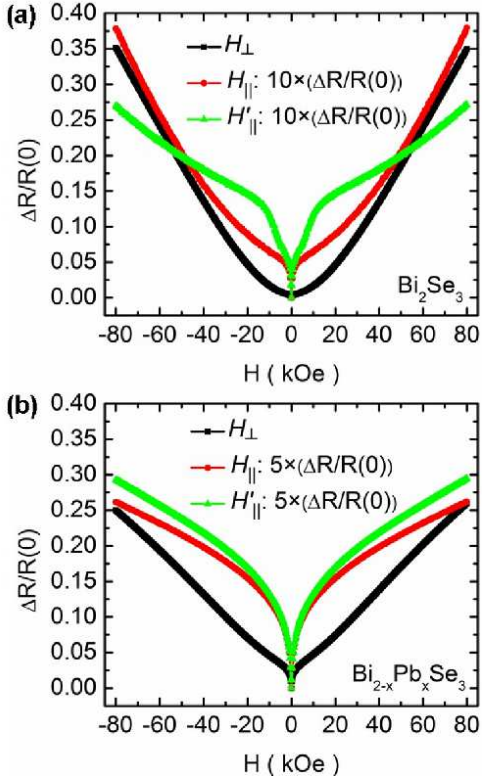}
\caption{Fractional change in magnetoresistance, $\Delta R(H)=(R(H)-R(0))/R(0)$, of the 45 nm thick (a) undoped Bi$_{2}$Se$_{3}$ and (b) doped Bi$_{2-x}$Pb$_{x}$Se$_{3}$ films. We used three different orientations of the magnetic field: $H_{\perp}$ denotes magnetic field perpendicular to the surface of the thin film, while $H_{\parallel}$ and $H_{\parallel}'$ denote an in plane magnetic field perpendicular to and parallel to the excitation current, respectively. For the latter two, the resistance ratio has been multiplied by the factor shown on the plot. All data are at 500 mK.}
\label{HighField}
\end{figure}

\begin{figure}
\includegraphics[width=.75\columnwidth]{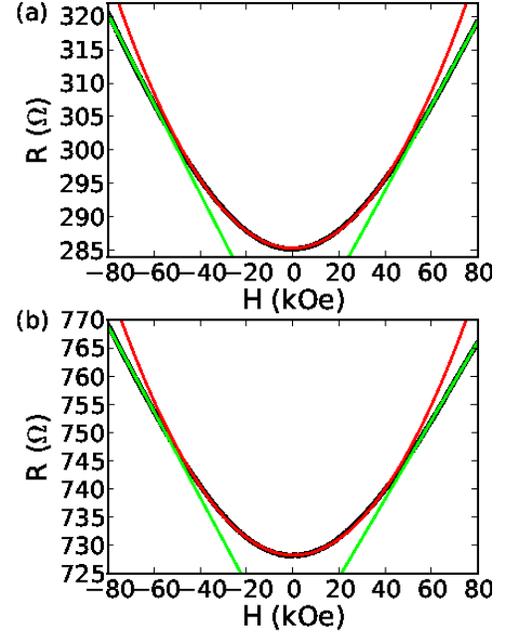}
\caption{Resistance vs Magnetic Field Plots at 300 K for the 45 QL Bi$_{2}$Se$_{3}$ and Bi$_{2-x}$Pb$_{x}$Se$_{3}$ Films in the Perpendicular Field.}
\label{MR}
\end{figure}

We have also measured the magnetoresistance for fields up to 80 kOe for three orientations of the magnetic field, as shown in Fig.~\ref{HighField}.
In the perpendicular field ($H_{\perp}$), the fractional MR change, $\Delta R/R_{0}=(R(H)-R(0))/R_{0}$, between 80 and 0 kOe is about 35\% for the Bi$_2$Se$_3$ film and 25\% for the Bi$_{2-x}$Pb$_{x}$Se$_{3}$ film. When H is parallel to the film, the fractional MR change is much smaller, between 2.7\% and 5.9\%.
Linear MR was observed above 10 kOe (30 kOe) when a perpendicular field was applied to the Bi$_{2-x}$Pb$_{x}$Se$_{3}$ (Bi$_2$Se$_3$) film as shown in Fig.~\ref{HighField}(b). Such a linear MR has been attributed to the quantum linear MR of the surface states~\cite{Tang2010}.  We note that even at room temperature, the MR above 45 kOe has a better fit with a linear rather than parabolic dependence on the magnetic field for both samples (see Fig. \ref{MR}). Upon closer inspection, the MR of the Bi$_2$Se$_3$ and Bi$_{2-x}$Pb$_x$Se$_3$ films shows sharp minima at zero field at low temperatures. These are better revealed in the Figure~\ref{WL}. 

In the following, we focus on the behavior at low temperatures and low magnetic fields, because that is the most relevant parameter regime for our discussion within the context of quantum theories of localization. Figures~\ref{EEI}(a) and (b) display the $T$ dependence of the conductance for zero magnetic field as well as for a 20 kOe in-plane field. The logarithmic behavior is suggestive of a weakly disordered 2D system where diffusive transport is determined by either weak localization or EEI. Figures ~\ref{EEI}(c) and~\ref{EEI}(d) show the magnetoconductance as a function of a perpendicular field $H_{\perp}$, while Figs. ~\ref{EEI}(e) and~\ref{EEI}(f) show the magnetoconductance for the two directions of an in-plane field ($H_{\parallel}$ and $H_{\parallel}'$) as defined in the caption of Fig.~\ref{HighField}. The observed magnetoconductance is negative, which is often referred to as weak {\em anti}localization. In all cases, the magnetoconductance peak disappears at temperatures above 20 K, suggesting quantum correction as its origin.

\section{Analysis}

\subsection{Weak localization for noninteracting electrons}

We first attempt to analyze these measurements using the standard results of weak localization (WL) theory for noninteracting electrons, and find that they are inconsistent with either 3D or 2D weak localization. The logarithmic temperature dependence of conductance rules out 3D behavior, because in 3D we expect $\Delta\sigma_{\rm WL}\propto T^{-p/2}$, where $p$ is defined by the $T$ dependence of the phase coherence time $\tau_{\phi}\propto T^{-p}$. The exponent $p$ is dependent on the source of the inelastic scattering which causes the phase decoherence. For example, $p=3$ for phonon scattering, and $p=2/3$ (3D) or $p=1$ (2D) for the electron-electron interaction. In 2D, on the other hand, the correction to the conductivity at zero field is given by~\cite{Hikami1980}
\begin{equation}\label{HLNT}
\Delta\sigma_{\rm WL}(T,H=0)=\frac{e^{2}}{2\pi^{2}\hbar}\:\alpha p\:\ln\left(\frac{T}{T_{0}}\right)
\end{equation}
where $\alpha$ is a constant depending on the relative strengths of the spin-orbit and spin-flip (magnetic) scatterings. In the limit of weak spin orbit and magnetic scattering, one obtains $\alpha=1$; in the limit of strong spin orbit scattering and weak magnetic scattering, one finds $\alpha=-1/2$; and when the magnetic scattering is strong, we have $\alpha=0$. The Hikami-Larkin-Nagaoka equation for magnetoresistance in perpendicular field is~\cite{Hikami1980},
\begin{eqnarray}
\label{HLN}
\Delta\sigma_{\rm WL}(H_\perp)&-&\Delta\sigma_{\rm WL}(0)=\alpha\:\frac{e^{2}}{2\pi^{2}\hbar}
\nonumber \\
&\times &\left[\psi\left(\frac{1}{2}+\frac{\hbar c}{4e\ell_{\phi}^{2}H_\perp}\right)-\ln\left(\frac{\hbar c}{4e\ell_{\phi}^{2}H_\perp}\right)\right]
\end{eqnarray}
where $\psi$ is the digamma function and $\ell_{\phi}$ is the inelastic scattering length. The positive slope of conductance vs. $\ln T$ implies that our samples are in the limit of weak spin-orbit scattering ($\alpha=1$). However, positive $\alpha$ implies {\em positive} magnetoconductance, which is qualitatively inconsistent with the observed anti-localization behavior.

\begin{figure}
\includegraphics[width=\columnwidth]{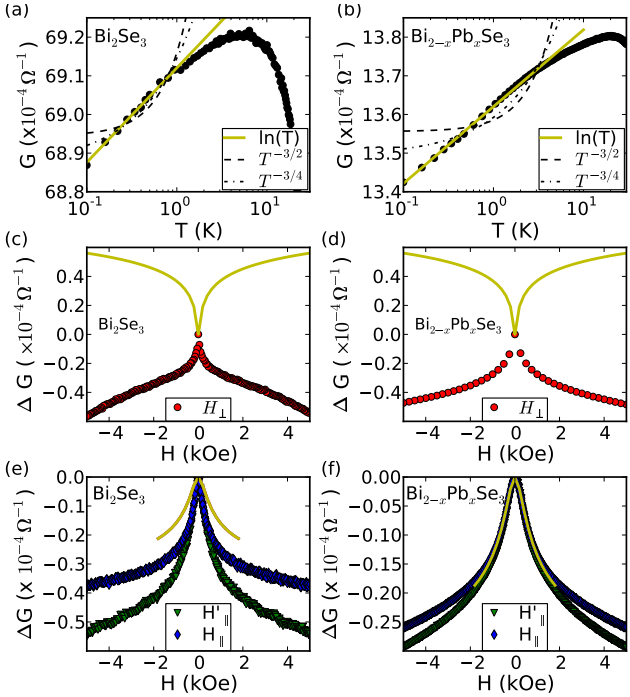}
\caption{Weak localization theory. (a-b) Temperature dependence of the conductance. Lines are fits to the weak localization theory in 2D (solid) and 3D (dashed and dash-dotted, corresponding to phonon and electron-electron scattering as the source of dephasing). The temperature dependence of both films suggests that our films are in the regime of weak spin-orbit scattering. However, in this regime, the theory predicts a positive magnetoconductance in perpendicular field. (c-d) Change in conductance, $\Delta G=\Delta G(H)-\Delta G(0)$. The theory predicts a localization effect, however we observe anti-localization. (e-f) Magnetoconductance in parallel fields ($H_{\parallel}$ denotes a field parallel to the film and perpendicular to the excitation current, while $H_{\parallel}'$ denotes a field parallel to both the film and the current). Lines are best fits to the theory of Maekawa and Fukuyama.\cite{Maekawa1981}}
\label{WL}
\end{figure}

Figure \ref{WL} shows explicit comparison of the experimental results to the weak localization theory. The lines in Figure \ref{WL} (c-d) show the weak localization theory (Eq. (2) in the main text) for $\alpha = 1$, as implied by the positive slopes of the data in Figure \ref{WL} (a-b). We have taken a typical value for the coherence length, $\ell_{\varphi}= 500$ nm. The lines in Figure \ref{WL} (e-f) are fits to the weak localization theory of Maekawa and Fukuyama\cite{Maekawa1981} in parallel field,
\begin{eqnarray}\label{Eqn-WL2}
\Delta\sigma_{\rm WL}(H)&-&\Delta\sigma_{\rm WL}(0)=-\frac{e^{2}}{4\pi^{2}}\left\{\ln\abs{1+2\left(\frac{y+z}{x-y}\right)\frac{\tau_{\phi}}{\tau}}{}\right. \nonumber\\
&-&\left. {}\frac{1}{\sqrt{1-\gamma}}\:\ln\abs{\frac{\frac{\tau}{\tau_{\phi}}+\left(\frac{y+z}{x-y}\right)\left(1+\sqrt{1-\gamma}\right)}{\frac{\tau}{\tau_{\phi}}+\left(\frac{y+z}{x-y}\right)\left(1-\sqrt{1-\gamma}\right)}}\right\}
\end{eqnarray}
where
\begin{eqnarray}
x&=&\frac{1}{\tau_{0}}\\
y&=&\frac{1}{\tau_{{\rm so},\perp}}\\
z&=&\frac{1}{\tau_{{\rm so},\parallel}}\\
\gamma&=&\left(\frac{g\mu_{B}B\tau}{\hbar}\left(\frac{x-y}{y+z}\right)\right)^{2}
\end{eqnarray}
$g$ is the Zeeman g-factor, $\tau_{0}$ refers to the scattering time from disorder, and $\tau_{{\rm so},i }$ refers to the spin-orbit scattering in the directions perpendicular and parallel to the film. The source of the magnetic field dependence is the Zeeman splitting only, and this effect vanishes for $g\rightarrow 0$. Roughly speaking, the factor $(y+z)/(x-y)$ is a measure of the comparative strengths of the spin orbit and elastic scattering. For example, if elastic scattering is strong (large $x$) compared to spin-orbit scattering, then this factor will be small. In the low-field limit, this equation gives $\Delta\sigma\propto B^{2}$, which is different from the observed sharp cusp. Nevertheless, as shown in Fig.~\ref{WL}(e) and (f), a very narrow parabola can be obtained for $(y+z)/(x-y) \approx 0.01$; i.e. weak spin orbit scattering ($x-y \ll y+z$). Although consistent with the temperature dependence, it is inconsistent with the expectation that the spin-orbit interaction in Bi$_{2}$Se$_{3}$ is strong.

\subsection{Electron-electron interaction}

Having shown that spin-orbit related WL alone cannot explain our data, we now examine the role of the dynamically screened EEI which also produces corrections to the conductivity. Again the logarithmic $T$ dependence rules out the 3D theory where we expect~\cite{Lee1985,Lee1982} $\Delta\sigma_{EEI}\propto\sqrt{T}$. On the other hand, the correction in 2D is
\begin{equation}\label{EEIT}
\Delta\sigma_{\rm EEI}(T)=\frac{e^{2}}{4\pi^{2}\hbar}\: \left(2-\frac{3}{2}\tilde{F}_{\sigma}\right)\ln\left(\frac{T}{T_{0}}\right)
\end{equation}
where $\tilde{F}_{\sigma}$ is a function of the average of the static screened Coulomb interaction over the Fermi surface, $F=\int d\hat{\Omega}\: v(q=2k_{F}\sin(\theta/2))/\int d\hat{\Omega}\: v(0)$. In 2D, $\tilde{F}_{\sigma}=-4+(8/F)(1+F/2)\ln(1+F/2)$. A non-zero magnetic field activates the large Zeeman splitting in Bi$_{2}$Se$_{3}$~\cite{Kohler1975}, introducing a further correction to the conductivity:~\cite{Lee1985}
\begin{equation}\label{GInteracting}
\Delta\sigma_{\rm EEI}(T,H)-\Delta\sigma_{\rm EEI}(T,0)=-\frac{e^{2}}{4\pi^{2}\hbar}\: \tilde{F}_{\sigma}\, g_{2} (T,H),
\end{equation}
where
\begin{equation}
g_{2}(T,H)=\int_{0}^{\infty}d\Omega\: \ln\left\lvert 1-\left(\frac{g\mu_B /k_B T}{\Omega}\right)^{2}\right\rvert\frac{d^{2}}{d\Omega^{2}}\frac{\Omega}{e^{\Omega}-1}.
\end{equation}
In Bi$_{2}$Se$_{3}$, the Zeeman $g$-factor is 23 when $H$ is in-plane, and 32 when $H$ is out of plane~\cite{Kohler1975}. When the Zeeman energy is much larger than $k_{B}T$, $g_{2}(H)\approx\ln(g\mu_{B} H/(1.3 k_{B} T))$. At 20 kOe (red data shown in Figs.~\ref{InteractingModel}(a) and (b)), this limit is satisfied for $T\ll 3$~K. The conductance is related to conductivity by a geometrical factor (e.g. length/width for a 2D surface) $\sim$1 for our samples.

\begin{figure}
\includegraphics[width=\columnwidth]{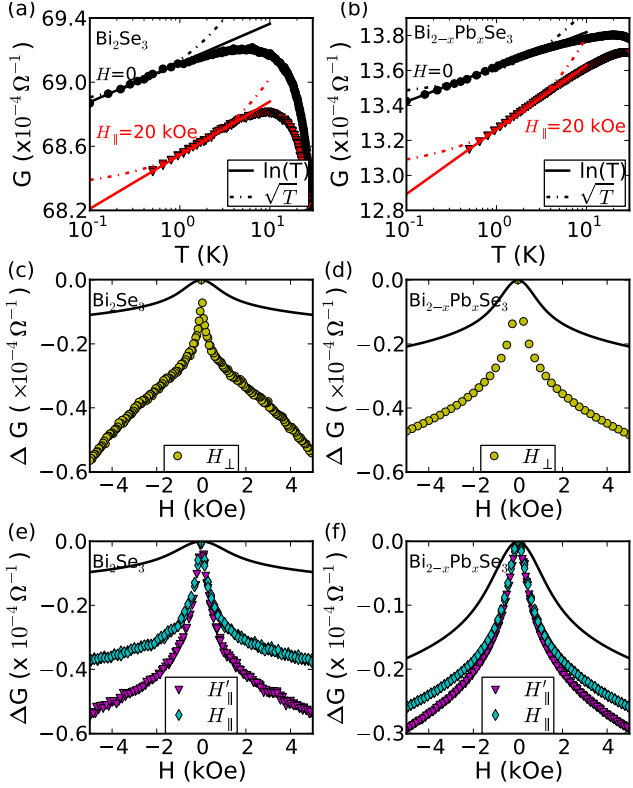}
\caption{Electron-electron interactions theory. (a-b) Temperature dependence of the conductance. Lines are fits to the 2D (solid) and 3D (dashed) theory of Lee and Ramakrishnan. The slope of the solid lines determines the screening parameter ($\tilde{F}_\sigma = 0.47$ and 0.75 for undoped and doped samples). Change in magnetoconductance in  (c-d) a perpendicular magnetic field and (e-f) parallel fields. Solid lines in (c-f) are fits to the electron-electron interaction theory using {\it no additional fitting parameters} besides those found in (a-b).}
\label{EEI}
\end{figure}

Figure \ref{EEI} shows a comparison of the data with the electron-electron interaction (EEI) theory. There are two fitting parameters, which are both obtained from the slopes of the lines in Figure \ref{EEI} (a-b). These are the screening parameter, $\tilde{F}_{\sigma}$, and a geometrical factor relating the conductance with the conductivity, $G=f\sigma$. In a wire with uniform current, $f$ is the ratio of cross-sectional area to length. We do not make this assumption, and instead use this factor as a parameter of the model. 
Using no further parameters, the EEI expressions correctly reproduce the {\em signs} of the $T$ and $H$ dependences, which already indicates the importance of EEI interactions. Although it does not capture the sharp peaks at zero field, the theory nicely reproduces all of the other qualitative features. 

\subsection{Combined weak localization and electron-electron interaction}

\begin{figure}
\includegraphics[width=\columnwidth]{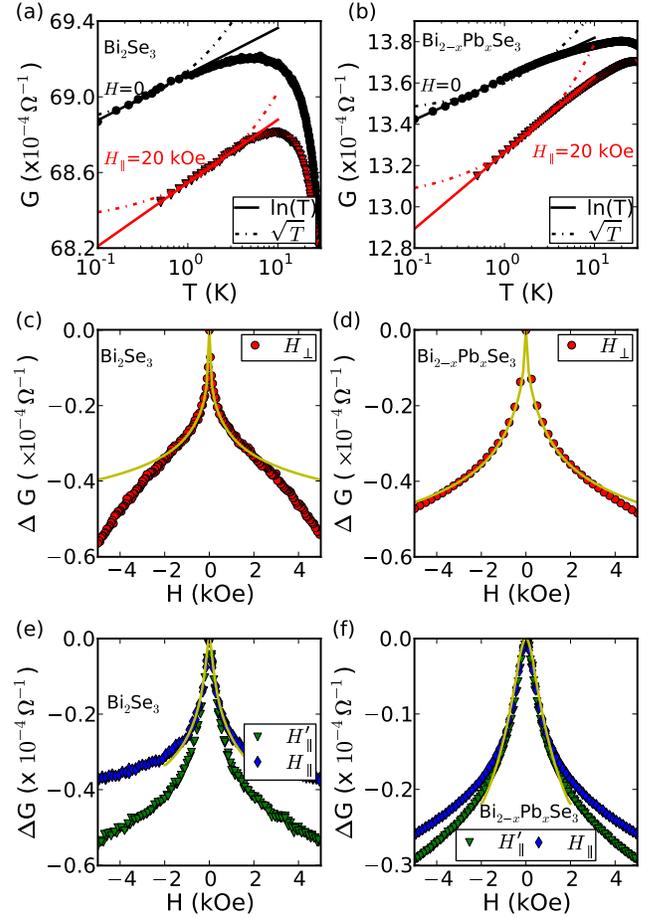}
\caption{Transport properties in low temperature and low magnetic field. 
The left columns are for the undoped Bi$_{2}$Se$_{3}$ film, while the doped Bi$_{2-x}$Pb$_{x}$Se$_{3}$ data is on the right. (a-b) Conductance vs $T$ for zero magnetic field (black circles) and for $H_{\parallel}=20$ kOe (red triangles), with fits to 2D theories (solid lines), and 3D electron-electron interaction (EEI) theory (dash-dotted lines).  (c-d) Magnetoconductance $\Delta G=G(H)-G(0)$ for fields perpendicular to the film. Solid lines are the result of a combined WL and EEI theory, as explained in the text. (e-f) Magnetoconductances for in plane fields $H_{\parallel}'$ and $H_{\parallel}$, parallel and perpendicular to the current direction. The solid lines are fits to the combined WL and EEI theory. The magnetoconductance data are all taken at 500 mK.}
\label{InteractingModel}
\end{figure}

We finally proceed to fit the experiment to the theory that contains corrections due to both weak localization and interaction. The temperature and H$_\perp$ dependences in Figure~\ref{InteractingModel}(a)-(d) are fitted to the combined model, using the approximation $\Delta\sigma(T,H_\perp)=\Delta\sigma_{\rm EEI}(T,H_\perp)+\Delta\sigma_{\rm WL}(T,H_\perp)$, with $\tilde{F}_{\sigma}$, $\alpha$ and $\ell_{\phi}$ taken as fitting parameters. (We take $p=1$, as appropriate for 2D systems at sufficiently low temperatures where EEI is the dominant mechanism for dephasing.) The data were fitted in the range  $-2$ kOe$<H<$2 kOe, which corresponds to the ``low-field'' region of the Bi$_{2}$Se$_{3}$ plot.  We find $\alpha=-0.31$ ($-0.35$),  $\ell_{\phi}=1.1$ ($0.64$)~$\mu$m, and $\tilde{F}_{\sigma}= 0.15$ ($0.35$) for Bi$_{2}$Se$_{3}$ (Bi$_{2-x}$Pb$_{x}$Se$_{3}$). The values for $\alpha$ are slightly different from the value of $-1/2$ expected from the strong spin-orbit interaction. The inelastic mean free paths are consistent with a previous measurement, $\ell_{\phi}=0.5\mu$m at 2 K~\cite{Peng2010}. 

The parameter $\tilde{F}_{\sigma}$, which incorporates the effect of interaction through $F$, can be estimated from first principles. While our sample thickness is much smaller than the inelastic scattering length, making it two dimensional insofar as the weak localization correction to conductivity  is concerned~\cite{Lee1985}, the thickness is much larger than the Thomas Fermi screening length ($k_{\rm TF}<1$ nm), suggesting that $F$ is to be estimated in a three-dimensional calculation. We will assume that to be the case.
For the carrier densities reported above, a straightforward calculation using the Thomas-Fermi approximation for the interaction gives $F=0.68$ ($\tilde{F}_{\sigma}=0.62$) and $0.72$ ($\tilde{F}_{\sigma}=0.65$) for the Bi$_{2}$Se$_{3}$ and Bi$_{2-x}$Pb$_{x}$Se$_{3}$ films.
For the calculation of $F$ we have assumed a three dimensional parabolic dispersion relation $p^2/2m^{*}$ where $m^{*}=0.15m_e$ ($m_e$ is the charge of an electron)~\cite{Kohler1974}. We have used the same dispersion for both samples, assuming that the doping has a negligible effect on the effective mass. (We note that if we had taken 2D screening with parabolic dispersion, we would get for $\tilde{F}$ 0.60 and 0.66, respectively. 
 A 2D calculation assuming a Dirac cone with Fermi velocity $v_F=5\times 10^5$ m/s yields a calculated $\tilde{F}=0.77$ independent of the Fermi wave vector.) 
The fitted values of $F (\tilde{F})$ are smaller, suggesting that we are overestimating the strength of the interaction. This discrepancy is not understood, but is not surprising in view of the crudeness of our model; specifically, we neglect: the strong anisotropy of the effective mass (which can change by a factor of 8 between parallel and perpendicular directions~\cite{Tichy}); doping dependence of the mass (which can change its value by as much as a factor of two~\cite{Tichy}); and corrections to the mass due to quantum confinement in one direction. 

We have also fit the data in parallel field, shown in Fig.~\ref{InteractingModel}(e-f). Since the theory does not allow for a difference between the two in-plane field directions, we have performed a simultaneous fit to both data sets. While the two data sets agree well at low fields, the disagreement at higher fields is not understood. In the limit of large spin-orbit scattering and low disorder, the predicted behavior cannot explain the sharp peak observed in the experiments. However, when the spin-orbit and disorder scattering times are comparable (roughly speaking, this occurs in the parameter range $(x+z)/(x-y)\sim 1 $), the weak localization model (Eqn.~\eqref{Eqn-WL2}) gives a good account of the observed behavior.

We note that there have been other experimental studies of thin films of TI materials~\cite{Chen2010b,He2010,Hirahara2010}. These do not identify the essential role of electron electron interactions. After the completion of this work we became aware of another article \cite{Liu2010} which studies transport in much thinner Bi$_2$Se$_3$ films (1-6 QLs) and interpret their results as indicating EEI as the origin of localization; their samples are not doped, however, and they also did not fit their data to EEI corrections discussed in the present article. 

\section{Conclusion}

In conclusion, we have studied two thin film samples of the Bi$_{2}$Se$_{3}$, one of which has surface states with Fermi energy located in the conduction band, while the other has a Fermi energy within the bulk band gap because of Pb doping. We find a 2D WL theory combined with EEI fits the transport data of both samples very well.  Note that the 2D WL model is relevant because the inelastic length in the these samples is much larger than the sample thickness,~\cite{Lee1985} and does not in itself indicate transport in a surface state. Although ARPES results indicate that the surface states should play a role in the conductance of the doped sample, it is not possible at this stage to single out the contribution from the surface states. What we can conclude with confidence is that the electron-electron interaction, which has been neglected in most previous analyses, plays an essential role in quantum corrections to transport in thin films of TI Bi$_2$Se$_3$ material.

We acknowledge financial support from the Penn State MRSEC under NSF grant DMR-0820404 and by DOE under grant no. DE- SC0005042.

$^*$ These authors contributed equally to the project.

\newpage


\begin{thebibliography}{99} 
\bibitem{Fu2007} L. Fu, C. L. Kane, and E. J. Mele, {\em Phys. Rev. Lett.}, {\bf 98}, 106803 (2007).
\bibitem{Moore2007} J. E. Moore and L. Balents, {\em Phys. Rev. B}, {\bf 75}, 121306 (2007).
\bibitem{Roy2009}R. Roy, {\em Phys. Rev. B}, {\bf 79}, 195321 (2009).
\bibitem{Qi2008}X. L. Qi, T. L. Hughes, S. C. Zhang, Phys. Rev. B 78, 195424 (2008).
\bibitem{Hsieh2009} D. Hsieh et al., Nature (London) 460, 1101 (2009).
\bibitem{Hasan2010}M. Z. Hasan and C. L. Kane, {\em Rev. Mod. Phys.}, {\bf 82}, 3045 (2010).
\bibitem{Zhang2009} H. J. Zhang, C. -X. Liu, X. L. Qi, X. Dai, Z. Fang, and S. -C. Zhang, Nature Phys. 5, 438 (2009).
\bibitem{Qi2010}X. -L. Qi and S. -C. Zhang, Phys. Today 63, No. 1, 33 (2010).
\bibitem{Checkelsky2009} J. G. Checkelsky et al., {\em Phys. Rev. Lett.}, {\bf 103}, 246601 (2009).
\bibitem{Peng2010} H. L. Peng et al., {Nature Materials}, {\bf 9}, 225 (2010).
\bibitem{Qu2010} D. -X. Qu et al., {\em Science}, {\bf 329}, 821 (2010).
\bibitem{Butch2010}N. P. Butch et al., {\em Phys. Rev. B}, {\bf 81}, 241301(R) (2010).
\bibitem{Cheng2010} P. Cheng et al., {\em Phys. Rev. Lett.}, {\bf 105}, 076801(2010).
\bibitem{Zhang2010} Y. Zhang et al., {\em Nature Phys.}, {\bf 6}, 584 (2010).
\bibitem{Lee1985} P. A. Lee, and T. V. Ramakrishnan, {\em Rev. Mod. Phys.}, {\bf 57}, 287 (1985).
\bibitem{Lee1982}P. A. Lee, and T. V. Ramakrishnan, {\em Phys. Rev. B}, {\bf 26}, 4009 (1982).
\bibitem{Kohler1975}H. K\"{o}hler, and E. W\"{o}chner, {\em Phys. Status Solidi B}, {\bf 67}, 665 (1975).
\bibitem{Zhang2010b}Y. Zhang, et al., {\em Appl. Phys. Lett.}, {\bf 97}, 194102 (2010).
\bibitem{Kong2010} D. S. Kong et al., {\em Nano Lett.}, {\bf 10}, 329 (2010).
\bibitem{Steinberg2010}H. Steinberg, {\em et al.}, {\em Nano Letters}, http://pubs.acs.org/doi/pdf/10.1021/nl1032183.
\bibitem{Tang2010} H. Tang et al., arXiv: 1003.6099v4.
\bibitem{Hikami1980} S. Hikami, A. I. Larkin, and Y. Nagaoka, {\em Prog. Theor. Phys.}, {\bf 63}, 707 (1980).
\bibitem{Maekawa1981}S. Maekawa and H. Fukuyama, {\em J. Phys. Soc. Jpn.}, {\bf 50}, 2516 (1981).
\bibitem{Kohler1974}H. K\"{o}hler and J. Kartmann, {\em Phys. Status Solidi. B},{\bf 63}, 171 (1974).
\bibitem{Tichy}L. Tich\'{y} and J. Hor\'{a}k, {\em Phys. Rev. B}, {\bf 19}, 1126 (1979).
\bibitem{Chen2010b} J. Chen et al., {\em Phys. Rev. Lett.}, {\bf 105}, 176602 (2010).
\bibitem{He2010} H. -T. He et al., arXiv: 1008.0141.
\bibitem{Hirahara2010}T. Hirahara et al. {\em Phys. Rev. B}, {\bf 82}, 155309 (2010).
\bibitem{Liu2010}M. Liu et al. arXiv: 1011.1055 (2010).
\end{thebibliography}
\end{document}